\documentclass[preprint2,numberedappendix]{emulateapj-rtx4}
\usepackage{graphicx}
\usepackage{times}
\usepackage{url}
\usepackage{bm}
\usepackage{natbib}
\graphicspath{{./fig/}{./png/}}
\usepackage{color}

\newcommand{\EQ}{\begin{equation}}
\newcommand{\EE}{\end{equation}}
\newcommand{\EQA}{\begin{eqnarray}}
\newcommand{\EEA}{\end{eqnarray}}
\newcommand{\brac}[1]{\langle #1 \rangle}
\newcommand{\pd}{\partial}

\newcommand{\ve}[1]{\boldsymbol{#1}}
\newcommand{\mean}[1]{\overline{#1}}

\newcommand{\urms}{u_{\rm rms}}

\newcommand{\Beq}{B_{\rm eq}}

\newcommand{\kef}{k_{\rm f}}

\newcommand{\tauto}{\tau_{\rm to}}
\newcommand{\chit}{\chi_{\rm SGS}}
\newcommand{\chitm}{\overline{\chi}_{\rm SGS}}

\newcommand{\Pm}{{\rm Pm}}
\newcommand{\Rm}{{\rm Rm}}
\newcommand{\Rey}{{\rm Re}}

\newcommand{\Pra}{{\rm Pr}}

\newcommand{\Rat}{{\rm Ra}_{\rm t}}

\newcommand{\Co}{{\rm Co}}

\def\onethird{{\textstyle{1\over3}}}
\def\onehalf{{\textstyle{1\over2}}}

\begin{document}

\title{Azimuthal dynamo wave in spherical shell convection}

\author{Elizabeth Cole$^{1}$, Petri J.\ K\"apyl\"a$^{1,2}$, Maarit J.\ Mantere$^{3,1}$ and Axel Brandenburg$^{2,4}$}
\affil{$^1$Physics Department, Gustaf H\"allstr\"omin katu 2a, PO Box 64,
FI-00014 University of Helsinki, Finland\\
$^2$NORDITA, KTH Royal Institute of Technology and Stockholm University,
Roslagstullsbacken 23, SE-10691 Stockholm, Sweden\\
$^3$Aalto University, Department of Information and Computer Science, 
PO Box 15400, FI-00076 Aalto, Finland \\
$^4$Department of Astronomy, AlbaNova University Center,
Stockholm University, SE-10691 Stockholm, Sweden}
\email{elizabeth.cole@helsinki.fi
($ $Revision: 1.108 $ $)
}

\begin{abstract}
  We report the finding of an azimuthal dynamo wave of a low-order ($m=1$) mode in direct
  numerical simulations (DNS) of turbulent convection in spherical shells. Such waves
  are predicted by mean field dynamo theory and have been obtained
  previously in mean-field models. 
  Observational results both from photometry and Doppler imaging have
  revealed persistent drifts of spots for several rapidly rotating
  stars, but, although an azimuthal dynamo wave has been proposed as
  a possible mechanism responsible for this behavior, 
  it has been judged as unlikely, as practical evidence for such waves
  from DNS has been lacking.
  The large-scale
  magnetic field in
  our DNS, which is due to self-consistent dynamo
  action, is dominated by a retrograde $m=1$ mode.
  Its pattern speed
  is nearly independent of latitude and
  does not reflect the speed of the differential rotation at any depth. 
  The extrema of magnetic $m=1$ structures coincide reasonably with
  the maxima of $m=2$ structures of the temperature.
  These results provide direct support for the observed drifts being
  due to an azimuthal dynamo wave.
\end{abstract}


\section{Introduction}
\label{intro}
The solar large-scale magnetic field is mostly axisymmetric and 
exhibits a dynamo wave propagating from mid-latitudes toward the
equator. The solar cycle is often explained in terms of $\alpha
\Omega$ dynamo models based on mean-field theory where the
poloidal field is regenerated via cyclonic turbulence ($\alpha$-effect)
and the toroidal field through differential rotation
($\Omega$-effect), see e.g.\ \cite{O03}.
The $\alpha$-effect is strongly anisotropic with more rapid rotation
\citep{Rue78} while its magnitude is less strongly
quenched than turbulent diffusivity \citep{KKB09a}.
At the same time, differential rotation is also quenched \citep[e.g.][]{KR99}, 
which enables non-axisymmetric modes to dominate.
Thus, in more rapidly rotating stars, the large-scale magnetic field
is expected to become more non-axisymmetric \citep{RWBMT90,MBBT95}.

Recent numerical simulations have reached a level of sophistication
where they have been able to produce oscillatory large-scale magnetic
fields \citep[e.g][]{GCS10,KKBMT10,BMBBT11,NBBMT13} and in some cases
 equatorward migration as in the Sun
\citep{KMB12a,KMCWB13,WKMB13}. Furthermore, as the rotation rate is
increased, non-axisymmetric large-scale fields are obtained
\citep{GD08,GDW12,KMCWB13}, as expected from mean-field dynamo theory. 

Observational results from photometry, spectroscopy and
spectropolarimetry show a similar trend for rapid rotators with high
levels of magnetic activity, manifested through extended high-latitude
starspots that have a predominantly non-axisymmetric longitudinal  
distribution \citep[e.g.][]{BT98,KMHI13}. The most often deduced
configuration consists of two active longitudes with
alternating levels of activity.  This is referred to as flip-flop
phenomenon since the original work of \citet{Jetsu1993} in the context
of phase jumps seen on the single giant star FK~Com.  With the
accumulation of observational data, it has become evident that the
flip-flopping does not occur periodically \citep[see
e.g.][]{Heidi07,HackmanFKCom13}. Moreover, in almost all cases the
phase behavior of the active longitude system shows disrupted linear
trends in the rotational frame of reference, i.e.\ the system is
usually not rotating with the same speed as the stellar surface. One
of the most prominent examples of this is the primary component of the
RS CVn binary system II Peg, where a drift pattern persistent over a
ten year epoch has been reported \citep{Hackea12,LMOPHHJS13}. Such drifts
are traditionally not explained by the presence of an azimuthal dynamo
wave, but by surface differential rotation causing the spots to
move with different speeds as their latitude changes, analogously to
the Sun. In some cases, such as FK~Com, the changing angular
velocity can clearly be related to changes in spot latitudes
\citep[see e.g.][]{Heidi07}. This picture, however, seems less evident
in II~Peg, in which no major changes in spot latitudes can be observed.
However, the magnetic structures move in the prograde direction with
respect to a rotating frame \citep{LMOPHHJS13}.

The idea that spots reflect the motion of the gas seems quite
straightforward, but there are various reasons why the pattern
speed associated with spots can be different from that of the gas.
Sunspots exhibit a prograde motion, which is often
associated with sunspots being anchored at some other depth
where it matches the local speed of the gas \citep[e.g.][]{PT98}.
However, in the Sun, magnetic tracers move usually faster
than the gas \citep{GDS03}, which can be explained as a
property of hexagon-like convection cells in the presence of
rotation \citep{Bu04,Bu07}, but it might also be related to the
near-surface shear layer in the Sun \citep{GK06,Br07}.
However, these are local considerations, so we should still
expect the pattern speed to reflect the equatorial acceleration
near the equator.
By contrast, in linear dynamo theory, a nonaxisymmetric
dynamo mode always rotates like a rigid body \citep{Ra86}.
Depending on model details and the sign of the $\alpha$ effect,
both prograde and retrograde rotation of the pattern is possible.
\cite{Ra86} discusses the so-called
westward drift of the Earth's magnetic field in the context of the geodynamo.
Rigidly rotating patterns also occur in the nonlinear regime \citep{RWBMT90}.
Thus, we should expect that dynamo patterns would not bear any information
about latitudinal differential rotation.

Here we report on simulations of rapidly rotating turbulent convection
that exhibit large-scale non-axisymmetric magnetic fields with
azimuthal dynamo waves.
We show that the pattern speed of nonaxisymmetric structures
is essentially constant, as expected from mean-field theory.

\section{The model}
\label{sec:model}

Our model is similar to that of \cite{KMB12a,KMCWB13}. We model a
shell in spherical polar coordinates, where $(r,\theta,\phi)$ denote
radius, colatitude, and longitude. Here we model a shell $r_0 \leq r
\leq R$, $\theta_0 \leq \theta \leq \pi-\theta_0$, and $0 \leq \phi
\leq \phi_0$, where $r_0=0.7\,R$, $\theta_0=\pi/12$, $\phi_0=2\pi$ and
$R$ is the radius of the star. We solve the compressible
hydromagnetic equations,
\begin{equation}
\frac{\pd \bm A}{\pd t} = {\bm u}\times{\bm B} - \mu_0\eta {\bm J},
\end{equation}
\begin{equation}
\frac{D \ln \rho}{Dt} = -\bm\nabla\cdot\bm{u},
\end{equation}
\begin{equation}
\frac{D\bm{u}}{Dt} = \bm{g} -2\bm\Omega_0\times\bm{u}+\frac{1}{\rho}
\left(\bm{J}\times\bm{B}-\bm\nabla p
+\bm\nabla \cdot 2\nu\rho\bm{\mathsf{S}}\right),
\end{equation}
\begin{equation}
T\frac{D s}{Dt} = \frac{1}{\rho}\left[-\bm\nabla \cdot
\left({\bm F^{\rm rad}}+ {\bm F^{\rm SGS}}\right) +
\mu_0 \eta {\bm J}^2\right] +2\nu \bm{\mathsf{S}}^2,
\label{equ:ss}
\end{equation}
where ${\bm A}$ is the magnetic vector potential, $\bm{u}$ is the
velocity, ${\bm B} =\bm\nabla\times{\bm A}$ is the magnetic field,
${\bm J} =\mu_0^{-1}\bm\nabla\times{\bm B}$ is the current density,
$\mu_0$ is the vacuum
permeability, $D/Dt = \pd/\pd t + \bm{u} \cdot \bm\nabla$ is the
advective time derivative, $\rho$ is the density, 
$\nu$ is the kinematic viscosity, $\eta$ is the magnetic diffusivity,
both assumed constant, ${\bm F}^{\rm rad}=-K\ve{\nabla} T$ and
${\bm F}^{\rm SGS} =-\chit \rho  T\ve{\nabla} s$
are the radiative and subgrid scale (SGS) heat fluxes, where $K$ is
the radiative heat conductivity and $\chit$ is the turbulent heat
conductivity, which represents the unresolved convective transport of
heat, $s$ is the specific entropy, $T$ is the temperature, and $p$ is
the pressure. The fluid obeys the ideal gas law with $p=(\gamma-1)\rho
e$, where $\gamma=c_{\rm P}/c_{\rm V}=5/3$ is the ratio of specific
heats at constant pressure and volume, respectively, and $e=c_{\rm V}
T$ is the specific internal energy. The rate of strain tensor
$\bm{\mathsf{S}}$ is given by
$\mathsf{S}_{ij}=\onehalf(u_{i;j}+u_{j;i}) -\onethird
\delta_{ij}\bm\nabla\cdot\bm{u}$, where the semicolons denote
covariant differentiation \citep{MTBM09}.

The gravitational acceleration is given by
$\bm{g}=-GM_\star\hat{\bm{r}}/r^2$, where $G$ is the gravitational constant,
$M_\star$ is the mass of the star (omitting the convection zone), and
$\hat{\bm{r}}$ is the unit vector in the radial direction.
The Rayleigh number in our simulation is much smaller than in real stars
due to the higher diffusivities. This implies higher energy fluxes and
thus larger Mach numbers \citep{BCNS05}. Furthermore, the rotation
vector $\bm\Omega_0$ is given by
$\bm\Omega_0=(\cos\theta,-\sin\theta,0)\Omega_0$.  To have realistic
Coriolis numbers, the angular velocity in the Coriolis force has to be
increased in proportion to the one third power of the increase of the
energy flux. The centrifugal force is omitted, as it would
otherwise be unrealistically large \cite[cf.][]{KMGBC11,KMCWB13}.

\subsection{Initial and boundary conditions}
\label{sec:initcond}
The initial state is isentropic and the hydrostatic temperature
gradient is given by $\pd T/\pd r=-g/[c_{\rm
    V}(\gamma-1)(n_{\rm ad}+1)]$, where $n_{\rm ad}=1.5$. We fix the
value of $\pd T/\pd r$ on the lower boundary. The density profile
follows from hydrostatic equilibrium. The heat conduction profile is
chosen so that radiative diffusion is responsible for supplying the
energy flux in the system, with $K$ decreasing proportional to
$r^{-15}$ \citep{KMCWB13} so that convection is responsible for the
majority of the energy transport. We use a weak, random Gaussian noise
small-scale seed magnetic field.

The radial and latitudinal boundaries are assumed to be impenetrable
and stress free, see Equations~(8) and (9) of \cite{KMCWB13}. For the
magnetic field we assume a radial field at the outer radial boundary
and perfect conductors at the latitudinal and lower radial boundaries,
see Equations~(10)--(12) of \cite{KMCWB13}. On the latitudinal
boundaries we assume that density and entropy have vanishing first
derivatives. On
the upper boundary we apply a black body condition
\begin{equation}
\sigma T^4  = -K\frac{\pd T}{\pd r} - \chit \rho T \frac{\pd s}{\pd r},
\label{eq:bbb}
\end{equation}
where $\sigma$ is a modified Stefan--Boltzmann constant
\citep[see][]{KMCWB13,BB13}.

\begin{deluxetable*}{cccccccccccccr}
\tabletypesize{\scriptsize}
\tablecaption{Summary of the runs.}
\tablecomments{grid size is $128\times256\times512$, $\Pm = 1$, $\xi = 0.02$,
$\mathcal{L}=3.8\cdot10^{-5}$, and 
$\tilde\sigma = \sigma R^2T_0^4/L_0 = 1.4\cdot 10^3$, where 
$T_0=T(r_0)$, for all runs.
Run B is referred to as Run E4 in \cite{KMCWB13}.
Furthermore, $\tilde{E}_{\rm mag}=E_{\rm mag}/E_{\rm kin}$, 
$\tilde{E}^{(0)}=E^{(0)}/E_{\rm mag}$ and $\tilde{E}^{(1)}=E^{(1)}/E_{\rm mag}$.}
\tablewidth{0pt}
\tablehead{
\colhead{Run} & 
\colhead{$\Pra_{\rm SGS}$} &\colhead{$\tilde\nu$} & \colhead{${\rm Ra}_{\rm t}$} & \colhead{$\Rey$} & \colhead{$\Rm$} & \colhead{$\Co$} & \colhead{$\tilde\Omega/\tilde\Omega_\odot$} & \colhead{$\tilde{E}_{\rm mag}$} & \colhead{$\tilde{E}^{(0)}$} & \colhead{$\tilde{E}^{(1)}$} & \colhead{$M$} & \colhead{$P$}
}
\startdata
A  & $3.5$ & $4.1\cdot10^{-5}$ & $1.7\cdot10^6$ & $26$ & $26$  &  $5.0$ & 2.7 & 0.312 & 0.166 & 0.047 & 0.834 & $-0.333$ \\ 
B  & $3.0$ & $3.5\cdot10^{-5}$ & $2.2\cdot10^6$ & $28$ & $28$  &  $8.1$ & 4.0 & 0.618 & 0.109 & 0.071 & 0.891 & 0.318 \\ 
C  & $3.0$ & $3.5\cdot10^{-5}$ & $2.6\cdot10^6$ & $24$ & $24$  & $15.5$ & 6.7 & 0.937 & 0.056 & 0.091 & 0.944 & 0.347 
\enddata
\label{tab:runs}
\end{deluxetable*}

\subsection{Dimensionless parameters}
As in \cite{KMCWB13}, we define our simulations by imposing the energy
flux at the bottom boundary, $F_{\rm b}=-(K \pd T/\pd r)|_{r=r_0}$
and the values of $\Omega_0$, $\nu$, $\eta$, and
$\chitm=\chit(r_{\rm m}=0.85\, R)$. The corresponding nondimensional
input parameters are the luminosity parameter $\mathcal{L} =
L_0/[\rho_0 (GM)^{3/2} R^{1/2}]$ where $L_0=4\pi r_0^2 F_{\rm b}$, 
and the (turbulent) fluid and
magnetic Prandtl numbers $\Pra=\nu/\chitm$ and $\Pm=\nu/\eta$, and the
non-dimensional viscosity $\tilde{\nu}=\nu/\sqrt{GMR}$.
The density stratification is controlled by the normalized
pressure scale height at the surface, $\xi = [(\gamma-1) c_{\rm
  V}T_1]/(GM/R)$.

Other useful diagnostic parameters are the fluid and magnetic Reynolds
numbers $\Rey=\urms/(\nu \kef)$ and $\Rm=\urms/(\eta \kef)$, where
$\kef=2\pi/\Delta r\approx21 R^{-1}$ is an estimate of the wavenumber
of the
largest eddies, and $\Delta r=R-r_0=0.3\,R$ is the thickness of the
layer. The Coriolis number is defined as $\Co=2\Omega_0\tauto$,
where $\tauto=(\urms \kef)^{-1}$ is the turnover time and
$\urms=\sqrt{(3/2)\brac{u_r^2+u_\theta^2}_{r\theta\phi\Delta t}}$ is
the rms velocity and the subscripts indicate averaging over $r$,
$\theta$, $\phi$, and a time interval $\Delta t$ of several
magnetic diffusion times during which the run is thermally relaxed.
The turbulent Rayleigh number $\Rat$ is quoted from the
thermally relaxed state of the runs
\begin{eqnarray}
\Rat\!=\!\frac{GM(\Delta r)^4}{\nu \chitm R^2} \bigg(-\frac{1}{c_{\rm P}}\frac{{\rm d}\brac{s}_{\theta \phi\Delta t}}{{\rm d}r} \bigg)_{r_{\rm m}}.
\label{equ:Ra}
\end{eqnarray}
We express the magnetic field in equipartition field strengths,
$\Beq(r)=\langle \mu_0 \rho \bm{u}^2 \rangle^{1/2}_{\theta\phi\Delta t}$,
where all three components of $\bm{u}$ are included. We average over
the $\phi$-coordinate to define mean quantities, denoted by an
overbar. Furthermore, we define magnetic end kinetic energies as
$E_{\rm mag}=\langle {\bm B}^2/2\mu_0 \rangle_{r\theta\phi\Delta t}$ and $E_{\rm
  kin}=\langle \rho {\bm u}^2/2 \rangle_{r\theta\phi\Delta t}$, and denote
the energies of the axisymmetric and $m=1$ modes of the magnetic field
as $E_{\rm mag}^{(0)}$ and $E_{\rm mag}^{(1)}$, respectively.
The simulations were performed with the {\sc Pencil
  Code}\footnote{http://pencil-code.googlecode.com/}.

\begin{figure*}[t]
\centering
\includegraphics[width=0.33\textwidth]{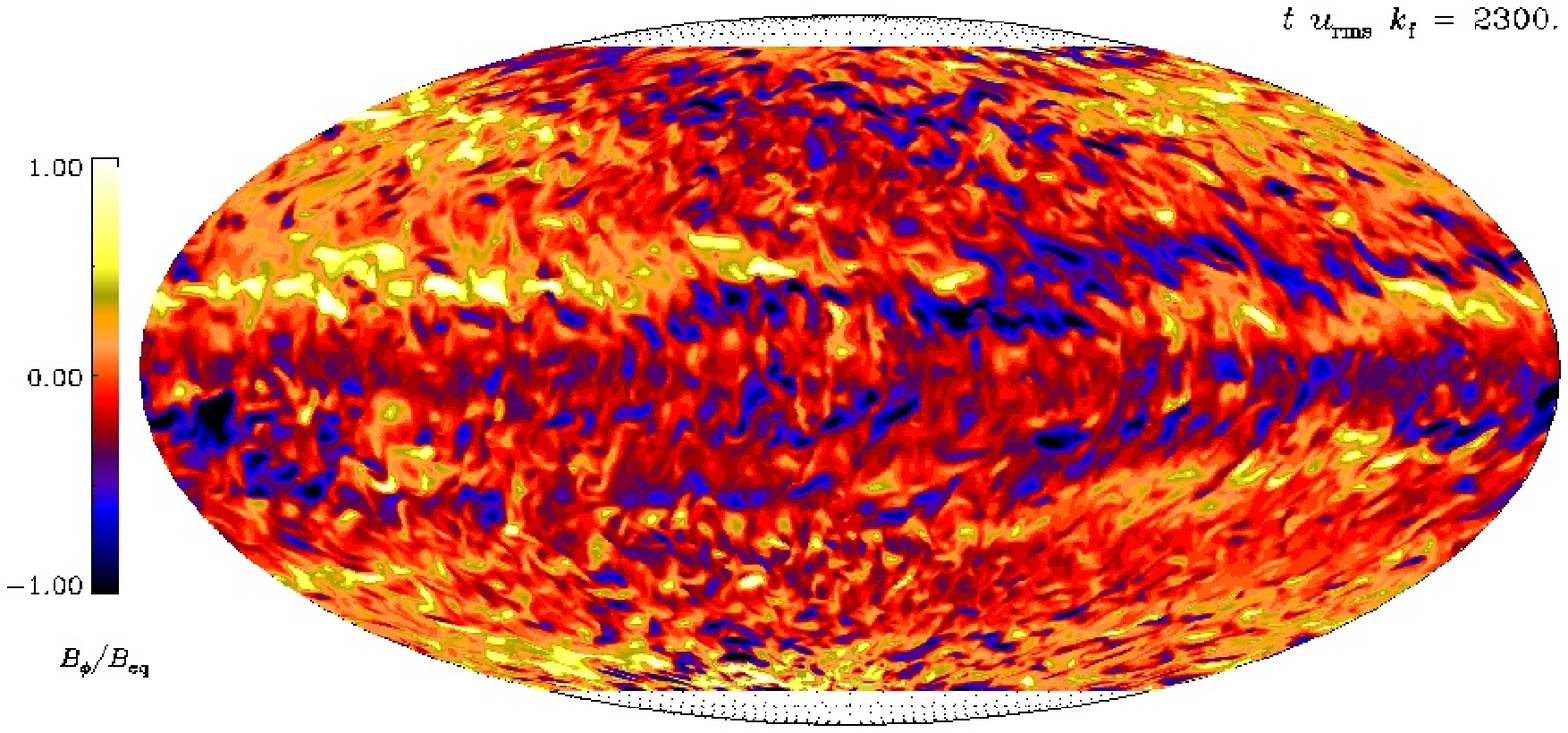}\includegraphics[width=0.33\textwidth]{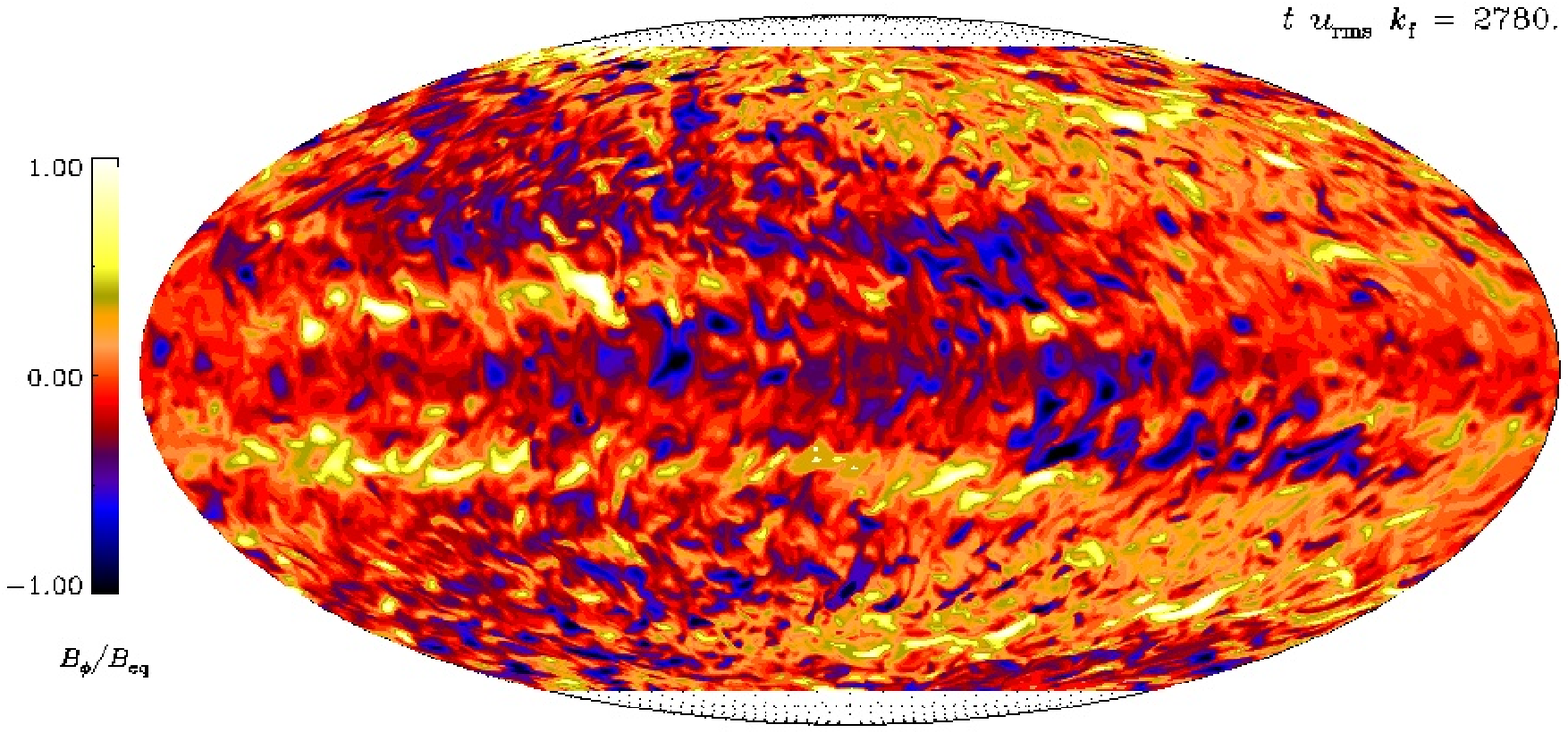}\includegraphics[width=0.33\textwidth]{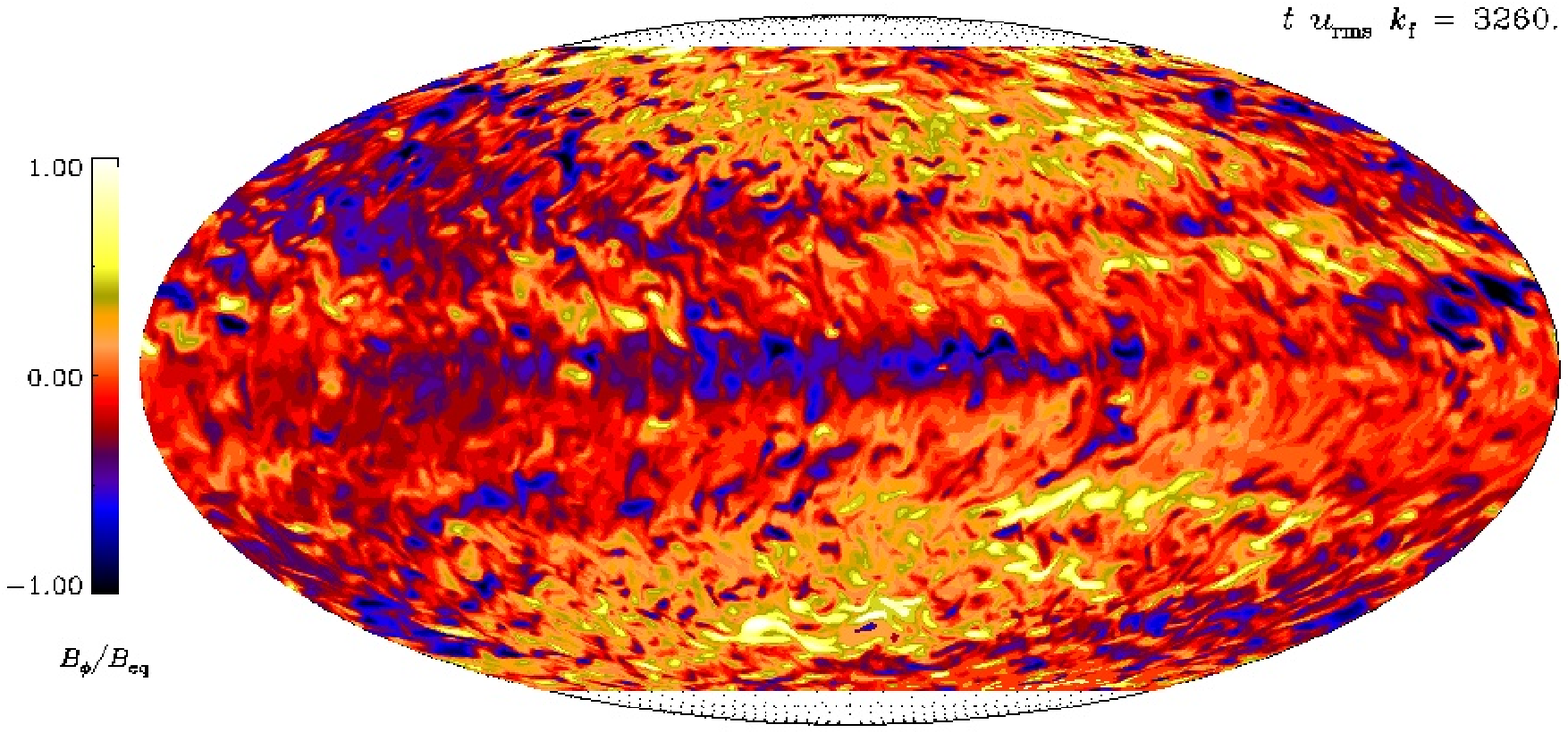}
\includegraphics[width=0.33\textwidth]{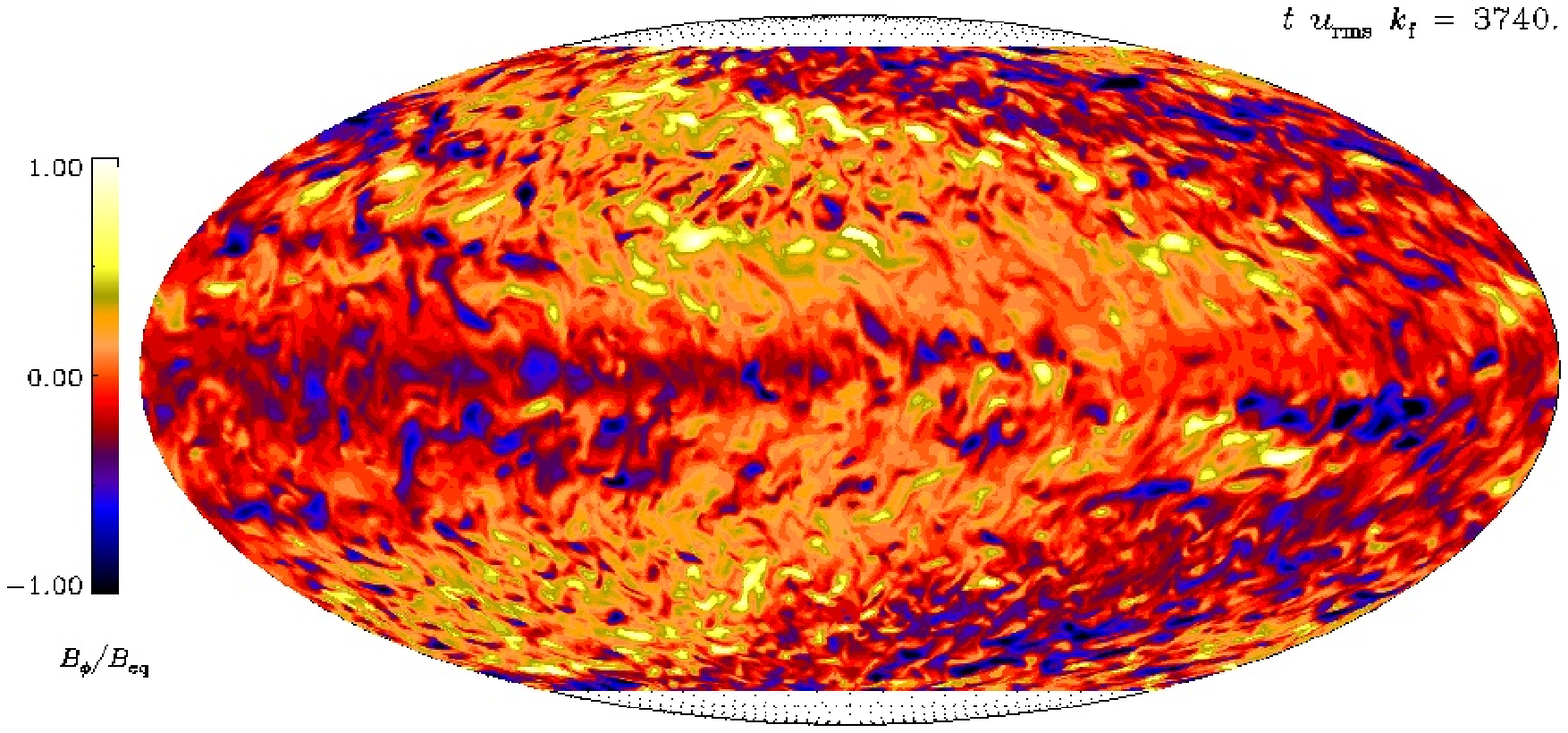}\includegraphics[width=0.33\textwidth]{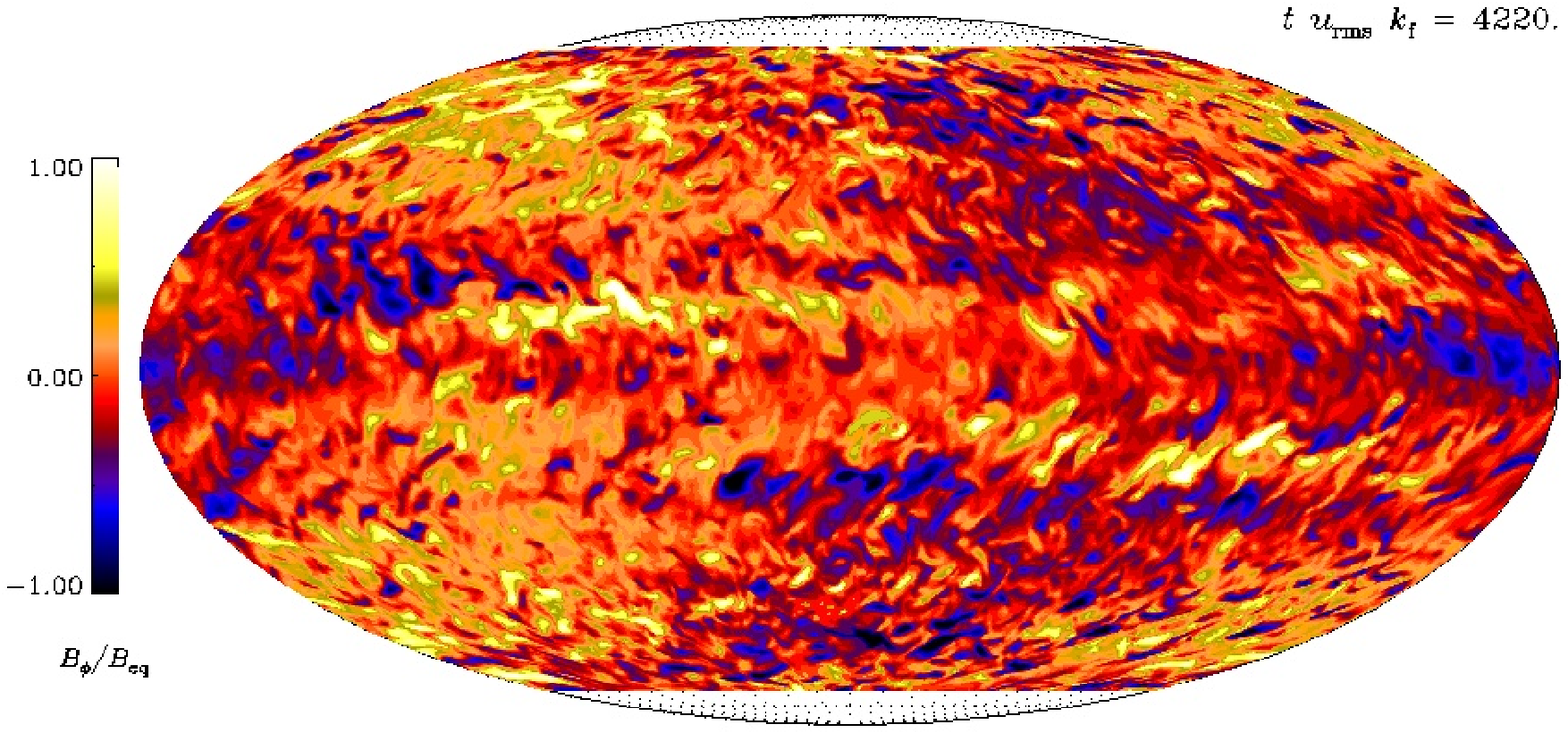}\includegraphics[width=0.33\textwidth]{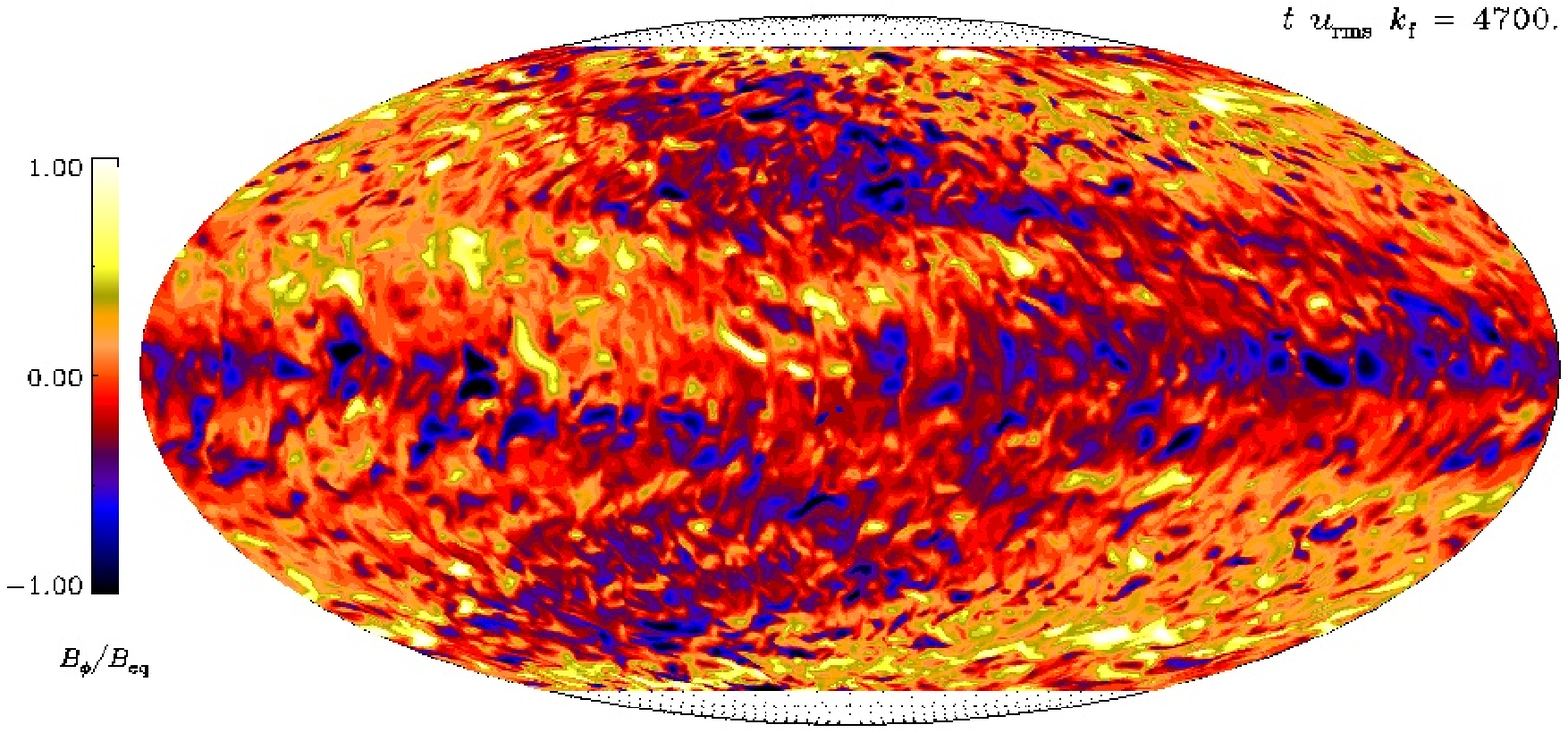}
\caption{
  Azimuthal magnetic field $B_\phi$, normalized by the equipartition 
  value $B_{\rm eq}$, near the surface of the star at
  $r=0.98R$ from Run~B for six times separated by $480\tauto$.}
\label{fig:pmoll_bb3}
\end{figure*}

\begin{figure*}[t]
\centering
\includegraphics[width=\textwidth]{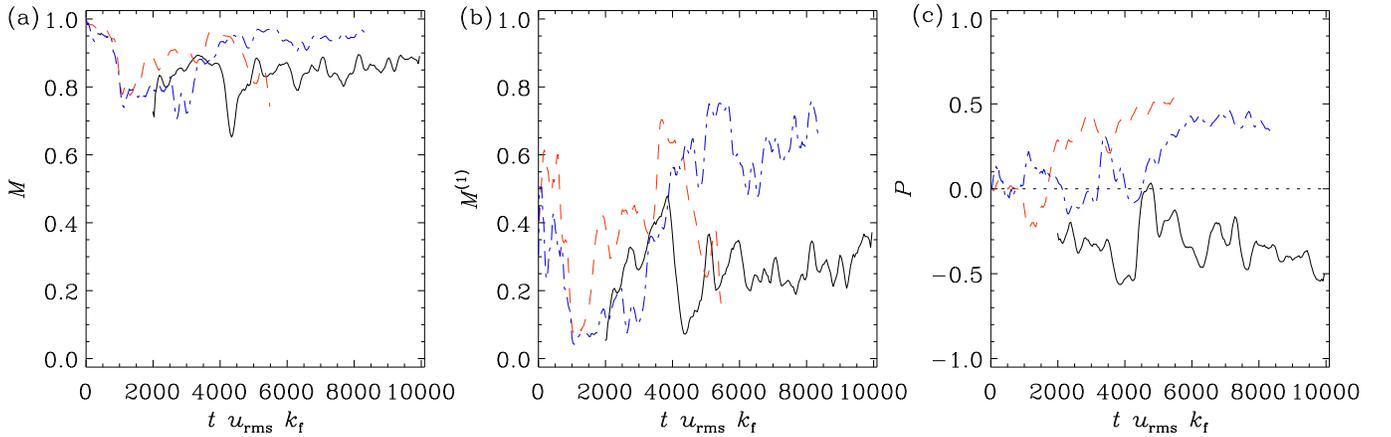}
\caption{Parameters $M$ ({\em left}), $M^{(1)}$ ({\em middle}), and
  $P$ ({\em right}) according to Equations~(\ref{equ:MM1}) and
  (\ref{equ:P}) from Runs A (black solid), B (red dashed), and C (blue
  dot-dashed). The data for Run~A does not start from zero due to a
  lack of diagnostics output from the early part of the run.}
\label{fig:pMP}
\end{figure*}

\subsection{Relation to stellar parameters}
We calibrate our model with solar parameters. However, due to the
compressible formulation of our model, using the real solar luminosity
would lead to prohibitively large Rayleigh numbers making the numerical
solutions 
infeasible. Thus we increase the luminosity in our model to
bring the dynamical and acoustic timescales closer to each other
\citep{KMCWB13}. The luminosity in our models is roughly $10^6$ times
greater than in the Sun. This means that the convective velocity is
roughly 100 times greater than in the Sun and we need to increase
$\Omega_0$ with the same factor to have the same rotational influence
as in the Sun.
We denote this solar-calibrated value by $\tilde\Omega_\odot$. In our
simulations the rotation rate is 2.7 to 6.7 times faster than this,
see Table~\ref{tab:runs}.

\section{Results}
\label{sec:results}
We discuss three simulations that can be interpreted
as representing the Sun at younger ages when it was rotating more
rapidly.
Runs~A, B, and C are respectively like Runs~B3m, B4m, and B5m
of \cite{KMB12a} where $\phi_0=\pi/2$, but are now with a full
$2\pi$ azimuthal extent and 20--40 per cent
higher viscosity and magnetic diffusivity.
Run~B is equivalent to Run~E4 of \cite{KMCWB13}.
We run the simulations from
the initial conditions discussed in Section~\ref{sec:initcond}.

\subsection{Nonaxisymmetric magnetic fields}

We find that in the early stages of the simulations an axisymmetric
oscillatory large-scale magnetic field grows first. This large-scale
component shows equatorward migration for the two highest rotation
rates \citep{KMB12a,KMCWB13}.  In the later stages, the dynamo mode
changes into a stable non-axisymmetric one, where the oscillations of
the axisymmetric part cease \citep[see Figure~17 of][]{KMCWB13}.

Figure~\ref{fig:pmoll_bb3} shows a sequence of snapshots of $B_\phi$
from Run~B near the surface at six times separated by $480\tauto$.
Strong magnetic fields occur as extended belts of toroidal field
near the equatorial region. These resemble the wreaths reported by
\cite{BBBMT10} and \cite{NBBMT13}, but instead of predominantly
axisymmetric structures, we now observe sign changes in longitude.
Strong magnetic fields are generated also at higher latitudes. These
structures appear to have a predominantly nonaxisymmetric distribution
with a large negative radial magnetic field on one side with a
positive counterpart on the other.

We Fourier filter the simulation data to extract the lowest order 
$m=0,1$ contributions to the magnetic field. We find that the energy
of the nonaxisymmetric $m=1$ mode is of the same order of magnitude as
the axisymmetric one in all runs (see Table~\ref{tab:runs}),
but observe a growth of the $m=1$ mode with respect to $m=0$ as
rotation increases. 
The lowest order modes constitute only roughly a fifth of the total
magnetic field energy, the rest being in still higher $(m>1)$ modes.

We quantify the non-axisymmetry of the magnetic field with the
quantities \citep[cf.][]{RWBMT90}
\begin{equation}
M = 1 - \frac{E_{\rm mag}^{(0)}}{E_{\rm mag}},\ \ \ M^{(1)} = 1 - \frac{E_{\rm mag}^{(0)}}{E_{\rm mag}^{(0)}+E_{\rm mag}^{(1)}}.
\label{equ:MM1}
\end{equation}
The random noise used as our initial
condition yields $M\approx M^{(1)}\approx1$.
A turbulent $m=0$ dynamo yields small values of $M^{(1)}=0$,
but, owing to contributions from random noise to high $m$ modes, $M=O(1)$,
while for a turbulent $m=1$ mode we have again $M\approx M^{(1)}\approx1$.
We show the time evolution of $M$ and $M^{(1)}$ in
Figures~\ref{fig:pMP}(a) and \ref{fig:pMP}(b).
We find that $M$ is close to unity in the saturated stages of our
runs. In earlier stages where the axisymmetric dynamo mode is more
prominent, the minimum values of $M$ are between $0.6$ and $0.7$,
whereas $M^{(1)}$ can be as low as $0.1$.
The ratio of the
nonaxisymmetric to axisymmetric field components is not completely
constant over time even in the saturated state, with variations of 
roughly 10 percent in comparison to the average values.
Furthermore, the larger the rotation rate, the closer is the
solution to a pure non-axisymmetric one.

We quantify the equatorial symmetry of the magnetic field
by the parity \citep{BKMMT89}
\begin{equation}
P=\frac{E^{\rm(S)}-E^{\rm(A)}}{E^{\rm(S)}+E^{\rm(A)}},
\label{equ:P}
\end{equation}
where $E^{\rm(S)}$ and $E^{\rm(A)}$ correspond to volume averaged energies
of the symmetric and antisymmetric parts of the magnetic field. The
extrema $P=1$ and $P=-1$ correspond to complete symmetry and complete
anti-symmetry with respect to the equator. A random initial field
produces $P=0$.
As apparent from Figure~\ref{fig:pMP}(c), there is mixed equatorial
symmetry at all times in all of the runs.
The lowest rotation case, Run~A, persistently shows preferentially
antisymmetric configuration, while Runs~B and C with increased
rotation evolve towards a symmetric configuration.

\subsection{Pattern speed of the $m=1$ structure}

Visual inspection of Figure~\ref{fig:pmoll_bb3} already reveals that
the large-scale non-axisymmetric structure is propagating in the
retrograde direction in the frame rotating with the star.
To analyze this drift quantitatively, we begin by using the
Fourier-filtered data at the surface of the star at $r/R=1$. We then
track the magnetic and temperature structures by following the extrema
of $B_r$ and $T$ of
the filtered (sinusoidal) signal.
We measure the azimuth $\varphi$ of the resulting $m=1$ structure,
and compute the pattern speed as $\Omega_{\rm pat}=d\varphi/dt$.
The resulting tracks of the magnetic extrema and corresponding
temperature maxima from Run~B are plotted in Figure~\ref{fig:pphase}
at different latitudes, together with the phase of the
differential rotation measured from the flow.
The signal in the magnetic field is well visible in both hemispheres,
while the one in temperature is more clear in the southern
hemisphere. For this reason, we perform the analysis only for the
southern hemisphere. 

From Figure~\ref{fig:pphase} it is evident that the nonaxisymmetric
structure is moving in the retrograde direction with nearly constant
speed. The pattern speed is considerably slower than the one expected
from advection by differential rotation at any latitude.
The $m=1$ structure completes an orbit in roughly
$2400\tauto$, whereas pure advection due to
differential rotation is typically 5--10 times faster.
From the analysis of the magnetic field and temperature at the
surface, it is evident that the nonaxisymmetric
structure rotates without being affected by differential rotation.

\subsection{Relation to local rotation rate}

Next we analyze the situation more thoroughly by computing the
rotation profile of the nonaxisymmetric mode $m=1$ from all the runs
as functions of latitude and depth, and compare it to the differential
rotation profiles, see Figure~\ref{fig:pOm_BB3}.
While the gas shows more differential rotation near the surface,
the $m=1$ pattern is essentially rigidly rotating; see
Figure~\ref{fig:pOm_BB3}(a).  In Figure~\ref{fig:pOm_BB3}(b) we show
the radial dependence of gas and pattern speeds.  The pattern speed is
retrograde by 0.09\%, showing a very small positive radial gradient
near the equator in the deeper part.  In all of these cases we find
that the magnetic structure is propagating in the retrograde
direction, see Figure~\ref{fig:pOm_BB3}(c). The normalized pattern
speed $\Delta \Omega_{\rm pat}= \Omega_{\rm pat}/\Omega_0$ is
monotonically decreasing as a function of rotation rate.
Our analysis, therefore, shows that the almost
rigid pattern speed $\Omega_{\rm pat}$ does not match the differential
rotation of the star at any depth or latitude. 

\begin{figure}[t]
\centering
\includegraphics[width=\columnwidth]{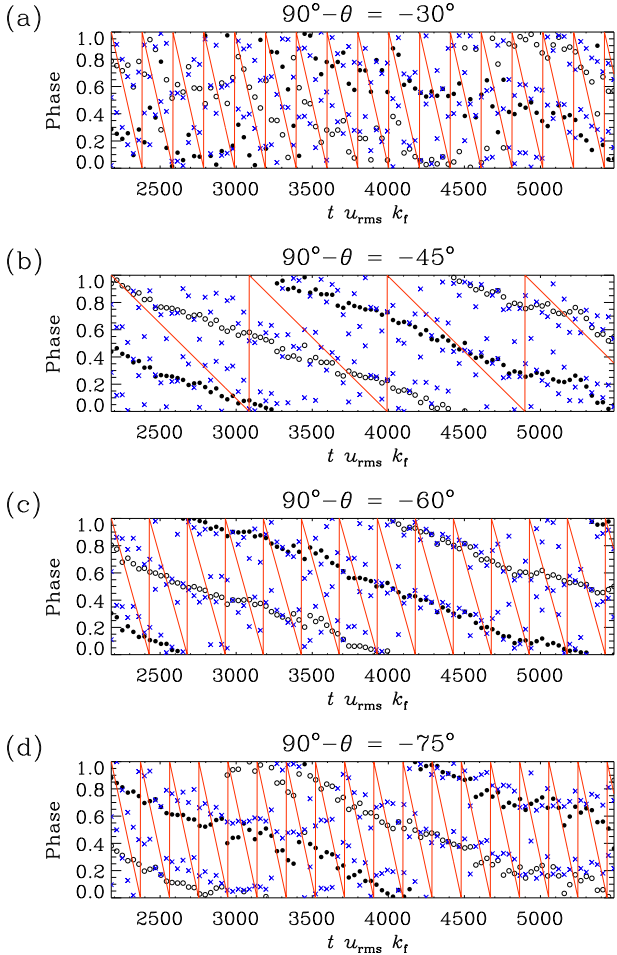}
\caption{Phase of the $m=1$ mode of $B_r$ (filled circles) and $-B_r$ (open circles) and the $m=2$ mode of $T$ (blue crosses) at the surface as a
  function time from Run B at four latitudes. Red lines denote phase based on pure
  advection due to differential rotation.}\label{fig:pphase}
\end{figure}

\begin{figure}[b]
\centering
\includegraphics[width=\columnwidth]{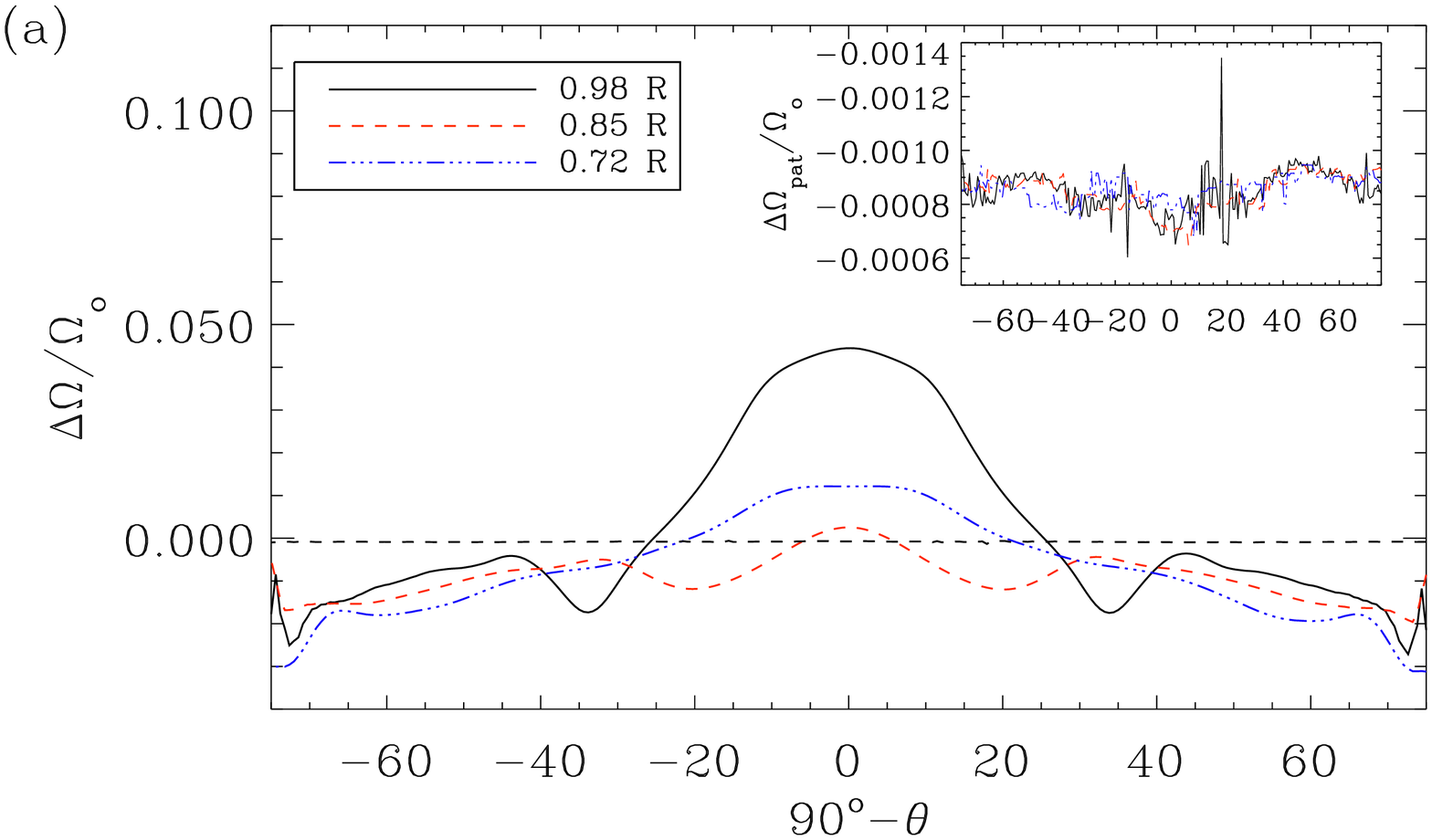}
\includegraphics[width=\columnwidth]{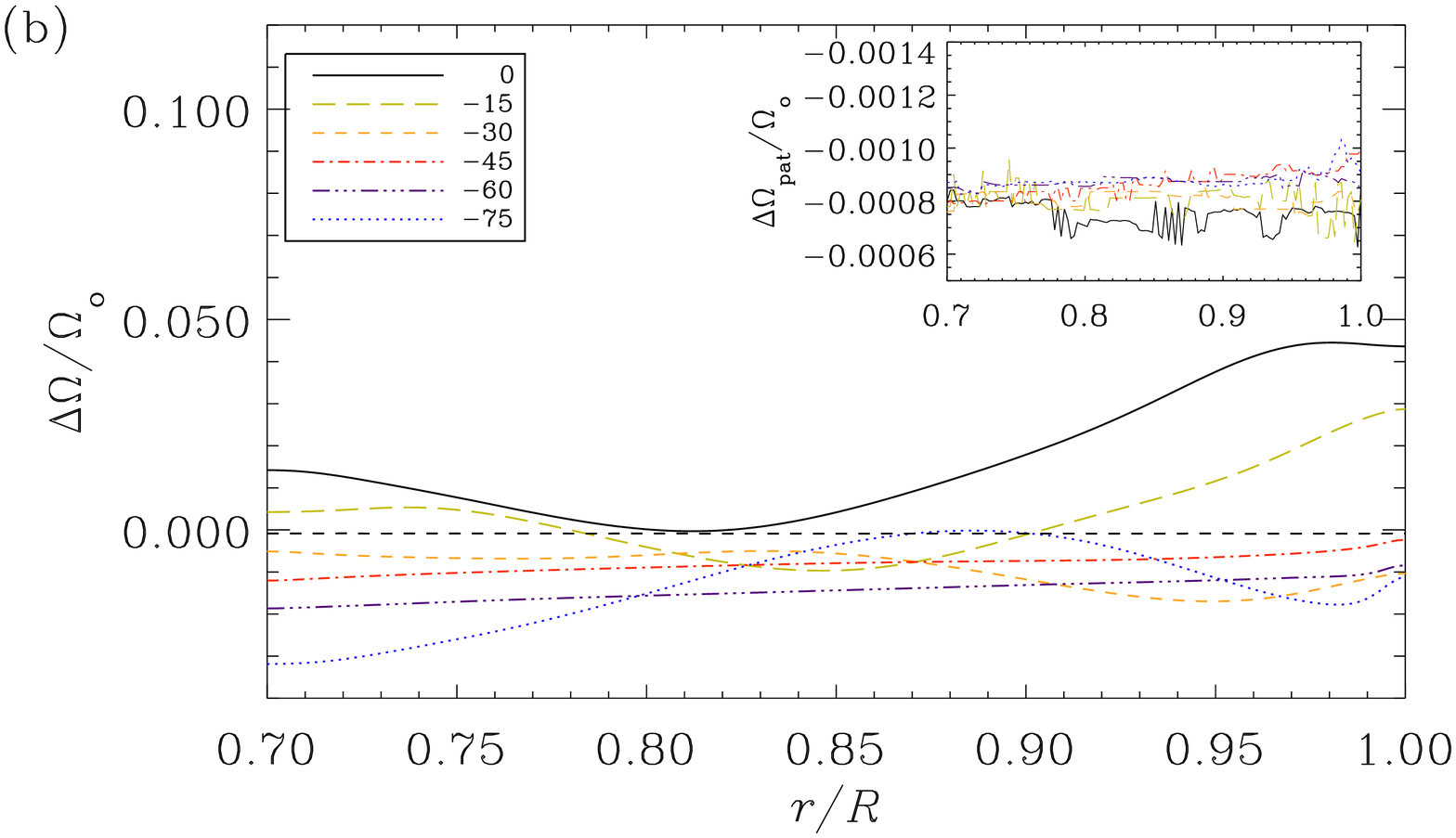}
\includegraphics[width=\columnwidth]{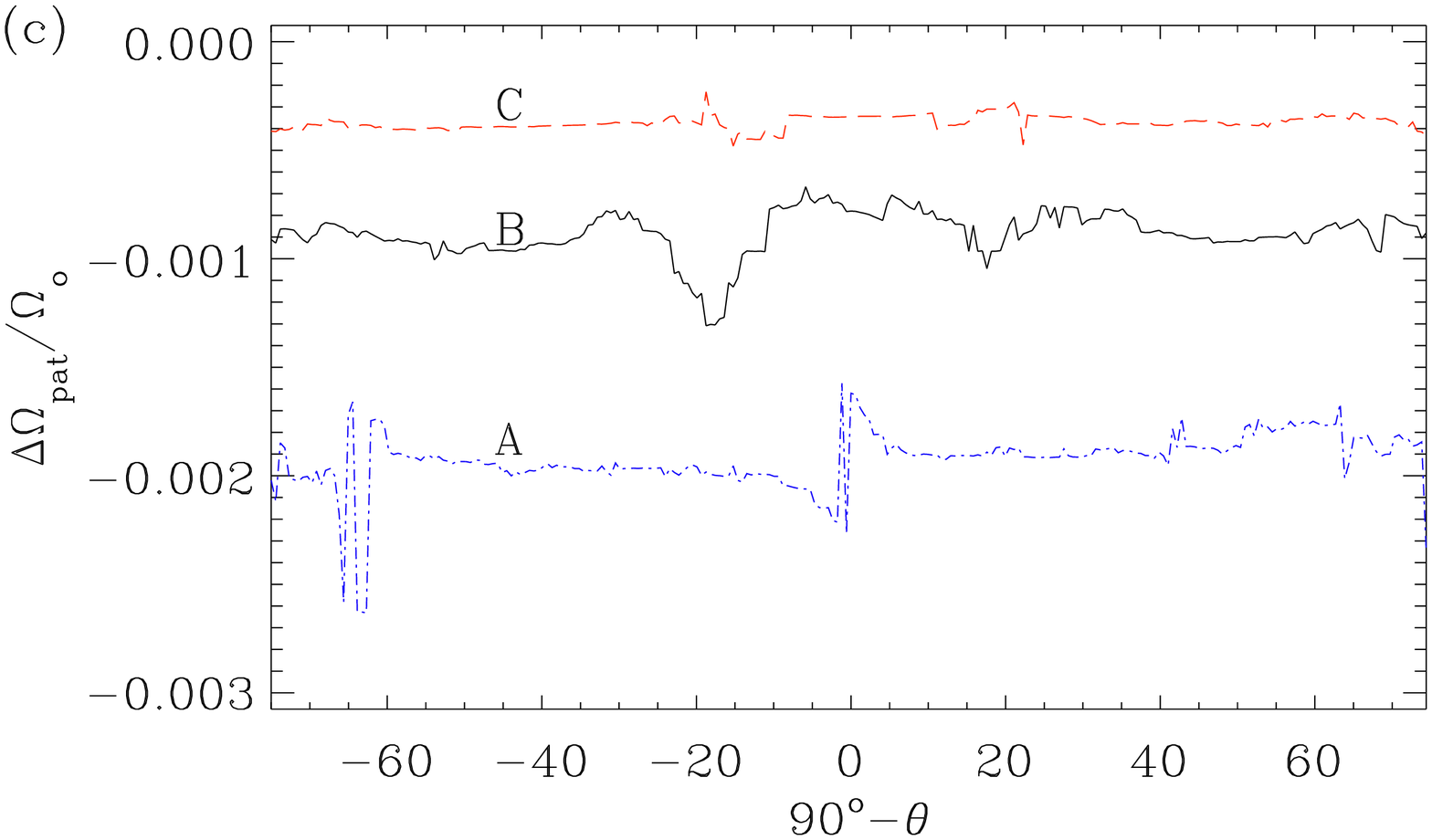}
\caption{{\em Top}: differential rotation 
  $\Delta\Omega/\Omega_0=\mean{u}_\phi/\Omega_0 r \sin \theta-1$
  from different depths and from the non-axisymmetric structure (black dashed
  horizontal line, enlarged in inset).
  {\em Middle}: angular velocity as a function of depth at select latitudes.
  {\em Bottom}: phase speed of the pattern, $\Omega_{\rm pat}= d\varphi/dt$
  from runs A, B, and C.}\label{fig:pOm_BB3}
\end{figure}

We also find a
corresponding signal in the temperature at high
latitudes; see Figure~\ref{fig:pphase}. We find that the extrema of
the $m=1$ contribution to $B_r$ match the maxima of the $m=2$
component of the temperature.
Temperature fluctuations $\Delta\mean{T}/\mean{T}$ of the $m=2$ mode
are 2--5 percent at high latitudes and 1--2 percent at mid- to low latitudes.
These fluctuations are largely independent of the normalized radial
component of the magnetic field, $\mean{B}_r/B_{\rm eq}$. This differs
from the results of Run~C1 from \citep{KMCWB13} where fluctuations between 15--20 percent were visible at high latitudes for the azimuthally averaged temperature components. 
 
\section{Conclusions}
\label{sec:conclusions}

We have studied azimuthal dynamo waves in three
dimensional simulations of convection-driven dynamos.
The wave moves rigidly and is generally
slower than the differentially rotating gas.
The drift cannot be accounted for by the differential rotation at any depth
in the simulations.
In the parameter regime investigated, only non-cyclic solutions with
retrograde patterns were found, and their speed is
decreasing with increasing rotation. In all cases investigated, the
non-axisymmetric pattern makes one orbit in the co-rotating frame in a
few thousand convective turnover times, 5--10 times slower than
expected if differential rotation of the fluid was the cause.

In active rapid rotators, the reported pattern speeds are of the
same order (in absolute terms) as reported in this study, see
\cite{LMOPHHJS13}. The deduced amounts of differential rotation for
these objects \citep[see e.g.][]{Henry1995,MCD05,Siwak10}, however, are much
weaker than the ones obtained in our numerical models. As a result, in
real objects the pattern speeds of non-axisymmetric structures is
only slightly smaller than or comparable to what is expected from advection
by differential rotation. Also, both prograde and retrograde drifts
have been observed \citep[see e.g.][]{BT98}, and commonly the patterns
are disrupted \citep[see e.g.][]{HackmanFKCom13}, the longest reported
drift so far being of the order of a decade \citep{BT98}.
The latitude-independent drift of the spot structure in II Peg
reported by \citet{LMOPHHJS13}, however, is not consistent with
latitude-dependent differential rotation, and more consistent with
the results presented here. In real objects, however, the dynamo
seems to operate in a regime where the azimuthal dynamo wave and
differential rotation have similar pattern speeds and these two
mechanisms compete with each other.

\acknowledgements
We thank Thomas Hackman for his comments on the manuscript.
The simulations were performed using the supercomputers hosted by CSC
-- IT Center for Science Ltd.\ in Espoo, Finland, who are administered
by the Finnish Ministry of Education. Financial support from the
Academy of Finland grants No.\ 136189, 140970 (PJK) and 218159, 141017 (MJM),
as well as the Swedish Research Council grants 621-2011-5076 and 2012-5797,
and the European Research Council under the AstroDyn Research Project
227952 are acknowledged.
The authors thank NORDITA for hospitality during their visits.

\bibliography{paper}

\end{document}